\documentclass{amsart}
\usepackage{amssymb}
\usepackage[footnotesize,bf]{caption}

\theoremstyle{definition}

\theoremstyle{remark}

\def\beq{\begin{eqnarray}}
\def\eeq{\end{eqnarray}}
\def\bsp{\begin{split}}
\def\esp{\end{split}}

\def\d{\mathrm{d}}

\newcommand{\mb}[1]{{\mathbb #1}}

\newcommand{\mbold}[1]{\mbox{\boldmath{\ensuremath{#1}}}}

\newcommand{\be}{\begin{equation}}
\newcommand{\ee}{\end{equation}}

\def \hbm #1 {\mbox{\boldmath{$\hat m^{(#1)}$}}}

\begin{document}
\hspace{9cm} NIKHEF/2007-016
\vspace{1cm}
\title{\textbf{Supergravity solutions with constant scalar invariants}}
\author{\textbf{A. Coley, A. Fuster and S. Hervik}}

\address{Department of Mathematics and Statistics,
Dalhousie University, Halifax, Nova Scotia, Canada B3H 3J5 (AC and SH); National Institute for Nuclear and High-Energy Physics
(NIKHEF), Kruislaan 409, 1098 SJ, Amsterdam, The Netherlands (AF); 
Faculty of Science and Technology,
 University of Stavanger,  N-4036 Stavanger, Norway (SH)}
\email{aac@mathstat.dal.ca; fuster@nikhef.nl; sigbjorn.hervik@uis.no}

\date{\today}

\maketitle

\begin{abstract}

We study a class of constant scalar invariant (CSI) spacetimes,
which belong to the higher-dimensional Kundt class,
that are solutions of supergravity. 
We review the known CSI supergravity solutions in this class and
we explicitly present a number
of new exact CSI supergravity solutions, some of which are
Einstein.

\end{abstract}

\noindent
[PACS: 04.20.Jb, 04.65.+e]

\vskip .1in

\section{Introduction}

A D-dimensional differentiable manifold of Lorentzian signature
for which all polynomial scalar invariants constructed from the
Riemann tensor and its covariant derivatives are constant is
called a constant scalar invariant (CSI) spacetime. The set of
spacetimes with vanishing scalar (curvature) invariants will be
denoted by VSI. The set of all locally homogeneous spacetimes will
be denoted by Hom. Clearly,  both VSI and homogeneous spacetimes
are CSI spacetimes; hence,  VSI $\subset$ CSI and Hom $\subset$
CSI.

Recently it was shown that the higher-dimensional VSI spacetimes
with fluxes and dilaton are solutions of type IIB supergravity,
and their supersymmetry properties \cite{VSISUG} were discussed
(also see \cite{ss1,ss2}). In this paper we shall study a
(sub)class of CSI spacetimes and determine whether they are
solutions of supergravity  (and discuss whether they can admit
supersymmetries). It is well known that $AdS_d \times S^{(D-d)}$
(in short $AdS\times S$) is an exact solution of supergravity (and
preserves the maximal number of supersymmetries). Of course, $AdS
\times S$ is an example of a CSI spacetime \cite{CSI}. There are a
number of other CSI spacetimes known to be solutions of
supergravity and admit supersymmetries; namely, there are
generalizations of $AdS \times S$ (for example, see
\cite{Gauntlett}), (generalizations of) the chiral null models
\cite{hortseyt}, and $AdS$ gyratons \cite{Caldarelli,FZ}.

We wish to find a class of CSI which are solutions of
supergravity and preserve supersymmetries. Clearly,
we seek as general a subclass as possible, but that will include
the simple generalizations of the $AdS\times S$ and $AdS$
gyratons. There are two possible approaches. In the {\em top-down}
approach, we can consider a subclass of known CSI spacetimes and
investigate whether they can be solutions of supergravity. For
example, we could consider product manifolds of the form $M\times
K$ (where, for example, $M$ is an Einstein space with negative
constant curvature and $K$ is a (compact) Einstein-Sasaki
spacetime). We could then use previous work to investigate whether
such spacetimes are solutions of supergravity and preserve
supersymmetries (cf. \cite{Kallosh}). Alternatively, we could use
a {\em bottom-up} approach in which we build CSI spacetimes using
known constructions \cite{CSI}. Although we are likely to find
less general CSI spacetimes of interest, the advantage of this
approach is that we can generate examples which by construction
will be  solutions of supergravity (provided that there are appropriate sources). 
We shall discuss both approaches below.

The set of all reducible CSI spacetimes that can be built from VSI
and Hom by (i) warped products (ii) fibered products, and (iii)
tensor sums \cite{CSI} are denoted by {CSI}$_R$. The set of
spacetimes for which there exists a frame with a null vector
$\ell$ such that all components of the Riemann tensor and its
covariants derivatives in this frame have the property that (i)
all positive boost weight components (with respect to $\ell$) are
zero and (ii) all zero boost weight components are constant are
denoted by {CSI}$_F$. Finally, those CSI spacetimes that belong to
the (higher-dimensional) Kundt class, the so-called Kundt CSI
spacetimes, are denoted by {CSI}$_K$. We note that by construction
{CSI}$_R$, and by definition {CSI}$_F$ and {CSI}$_K$, are at {\it
most} of Riemann type $II$ (i.e., of type $II$, $III$, $N$ or $O$
\cite{class}). In \cite{CSI} it was conjectured that  if a
spacetime is {CSI}, then the spacetime is either locally
homogeneous or belongs to the higher-dimensional Kundt CSI class
(i.e., {CSI}$_K$), and if a spacetime is {CSI}, then it can be
constructed from locally homogeneous spaces and VSI spacetimes\footnote{All of these conjectures have been proven in three dimensions \cite{3dCSI}.}.
This construction can be done by means of fibering, warping and
tensor sums (i.e., {CSI}$_R$). Thus, it is plausible that for CSI
spacetimes that are not locally homogeneous, the Weyl type is
$II$, $III$, $N$ or $O$, and that all boost weight zero terms are
constant (i.e., {CSI}$_F$).

\subsection{Higher-dimensional Kundt spacetimes}

The generalized D-dimensional Kundt CSI$_K$ metric can be written
\cite{CSI} \beq \d s^2=2\d u\left[\d v +H(v,u,x^k)\d
u+W_{i}(v,u,x^k)\d x^i\right]+g^{\perp}_{ij}(x^k)\d x^i\d x^j,
\label{HKundt}\eeq where the metric functions $H$ and $W_ {i}$ are
given by
\beq
W_{i}(v,u,x^k)&=& v{W}_{
i}^{(1)}(u,x^k)+{W}_{
i}^{(0)}(u,x^k),\label{HKa}\\
H(v,u,x^k)&=& {v^2}\tilde{\sigma}+v{H}^{(1)}
(u,x^k)+{H}^{(0)}(u,x^k), \label{HKb} \\
\tilde{\sigma} &=& \frac 18\left(4\sigma+W^{(1)i}W^{(1)}_i\right),
\label{sigma} \eeq (and are subject to further differential
constraints) and the transverse metric (where $\d
S_{hom}^2=g^{\perp}_{ij}\d x^i\d x^j$ is a locally homogeneous
space) satisfies the Einstein equations (where $i,j = 2, ...,
D-2$).

VSI spacetimes, with metric $\d s_{VSI}^2$, are of the form
(\ref{HKundt})  with flat transverse metric (i.e.,
$g^{\perp}_{ij}={\delta}_{ij}$) and the constant $\sigma$ in
(\ref{sigma}) is zero (and where the metric functions $H$ and
$W_{i}$ satisfy additional conditions) \cite{CFHP}.

For a $CSI_K$ spacetime the zero boost weight components of the
Riemann tensor, $R_{ijmn} = {R}^{\perp}_{i jm n} $, where ${
R}^{\perp}$ denotes the Riemann tensor components of the
transverse metric, are all constant \cite{CSI}. In general, the
Weyl and Ricci types of the CSI$_K$ spacetime is $II$
\cite{class}. A CSI$_K$ spacetime is of Ricci type $III$ when
$R_{01}={R}^{\perp}_{ij}=0$, and is of Ricci type $N$ if, in
addition,  $R_{1i}=0$ (Ricci type $O$ is vacuum).

The higher-dimensional Kundt  metric (\ref{HKundt}) possesses a
null vector field $\ell \equiv
\partial/\partial{v}$ which is geodesic, non-expanding, shear-free
and non-twisting \cite{Higher}. The aligned, repeated, null vector
$\ell$  is a null Killing vector (KV) in a $CSI_K$ spacetime if
and only if $H_{,v}=0$ and $W_{i,v}=0$, whence the metric no
longer has any $v$ dependence, and $\ell$ is, in fact, a
covariantly constant null vector (CCNV) \cite{CFHP}. In this case
the resulting spacetime is a product manifold with a CCNV-VSI
Lorentzian piece of Ricci and Weyl type $III$ and a locally
homogeneous transverse Riemannian space of Ricci and Weyl type
$II$ (in general).

\section{Analysis}

\subsection{Top-down approach}

It is well known that $AdS_d \times S^{(D-d)}$ is an exact
solution of supergravity (for certain values of $(D,d)$ and for
particular ratios of the radii of curvature of the two space
forms; in particular, $d=5,D=10$, $AdS_5 \times S^{5}$). Suppose
the more general $D$-dimensional product spacetime $M_d \times
K^{(D-d)}$  (in brief $M \times K$) is considered, where $M$ is an
Einstein space and $K$ is compact (e.g., a sphere, or a compact
Einstein space). We can ask: What are the most general forms for $M$
and $K$ such that the resulting product spacetime is an exact
solution of some supergravity theory (for a particular dimension,
and any particular fluxes)?  In particular, for $(D,d)=(11,4),(11,7)$ and $(5,5)$ it is sufficient that $M$ and $K$ are Einstein. Since  $M \times K$ is a Freund-Rubin
background, then if $M$ is any Lorentzian Einstein manifold and
$K$ is any Riemannian Einstein manifold (with the same ratio of
the radii of curvature as in the $AdS \times S$ case), then $M
\times K$ is a solution of some supergravity theory (not worrying
about whether the solution preserves any supersymmetry at the
moment). The fluxes are given purely in terms of the volume forms
of the relevant factor(s).  In general, the supergravity equations
of motion force $M$ to have negative scalar curvature and $K$ to
have positive scalar curvature (in order to be able to take $K$ to
be hyperbolic space exotic supergravity theories need to be considered).

$AdS \times S$ is an example of a spacetime manifold in which all
curvature invariants (including differential invariants) are
constant. Indeed, it is even a Kundt spacetime; i.e., it is a
CSI$_K$ spacetime. There are many examples of CSI spacetimes in
the Freund-Rubin $M \times K$ supergravity set. $K$ could be a
homogeneous space or a space of constant curvature. 

The question then is whether these CSI solutions preserve any
supersymmetry. Suppose that $M \times K$ is a Freund-Rubin
background.  The condition for preservation of supersymmetry
demands that $M$ and $K$ admit Killing spinors (real for $K$
(Riemannian) and imaginary for $M$ (Lorentzian)).  For $K$, the
existence of such spinors implies that $K$ is an Einstein space,
whereas for $M$ it must be imposed as an additional assumption. The analysis
therefore reduces to determining which Riemannian and Lorentzian
local metrics admit Killing spinors. The Riemannian case is well
understood (at least in low dimension -- for Freund-Rubin one
needs $d < 8$), but the Lorentzian case is still largely open. For
example,  the amount of supersymmetry preserved in supergravity
solutions which are the product of an anti-de Sitter space with an
Einstein space was studied in \cite{Acharya}. We note that there
are many homogeneous (CSI) examples of Freund-Rubin backgrounds.

More general results are possible. For example, suppose that $M
\times K$ is a Freund - Rubin manifold in which $M$ and the
compact $K$ are both Einstein spaces (and the signs and magnitudes
of the cosmological constants are appropriately arranged), then if
$M$ admits a conformal Killing vector (spacelike, and a negative
cosmological constant) then $M \times K$ is an exact solution of
supergravity \cite{Guven}. In a more general sense, any CSI
spacetime of the form $M \times K$ for which the Ricci tensor is
of type $N$ \cite{CSI} can be a solution of supergravity if appropriate sources exist. In
addition, in general if such a CSI spacetime admits a Killing
spinor, it would then give rise to a null (or timelike) Killing
vector (e.g., it would be a CCNV spacetime). These spacetimes
would then be of interest if there exist source fields that
support the supergravity solution and are consistent with the
supersymmetry.

\subsection{Bottom-up approach}

We want to construct as general a subclass of CSI spacetimes as
possible which are generalizations of $AdS\times S$ or $AdS$
gyratons, perhaps restricting attention to CCNV and Ricci type $N$
spacetimes. We shall start with a seed solution and then attempt
to build up an appropriate solution. In particular, we shall build
subsets of CSI$_K$ and CSI$_F$, by constructing CSI$_R$ spacetimes
using a VSI seed and locally homogeneous (Einstein) spaces.
Generalizations of $AdS\times S$ or $AdS$ gyratons can be
constructed in this way.

We construct a class of CSI$_R$ spacetimes from VSI and locally
homogeneous spacetimes as follows \cite{CSI}. We begin with a
general $d$-dimensional VSI spacetime, with metric $\d s_{\text{VSI}}^2$ given
by (\ref{HKundt}). We then warp this metric with warp factor
$\omega^2$. If the VSI metric is Ricci flat (i.e., a $d$-dimensional
vacuum solution; this implies certain differential conditions on
$H$ and $W_i$), and $\omega= l/z$ (where $l$ is constant:
curvature radius of $AdS$), then $\omega^2 \d s_{\text{VSI}}^2$ is an
Einstein space with $\lambda=-(d-1)/l^2$ and therefore a
$d$-dimensional vacuum solution with $\Lambda=-(d-1)(d-2)/(2l^2)$
(but where $H$ and $W_i$ satisfy now different equations). 
Their Ricci type is $II$ (and not lower). On the
other hand, since $\omega^2 \d s_{\text{VSI}}^2$ is conformal to
$\d s_{\text{VSI}}^2$, their Weyl type is the same (III at most). By
construction, all of these metrics have the same (constant)
curvature invariants as $AdS$. Indeed, the spacetimes constructed
from a CCNV-VSI (where the metric functions have no
$v$-dependence; for example, the $AdS$ gyraton) have a null
Killing vector, which makes them attractive from a supersymmetry
point of view. Note, however, that these spacetimes are not
necessarily CCNV themselves. It is unlikely (although possible for special cases) that spacetimes
constructed from a non-CCNV VSI will have
any null or timelike Killing vector. \\

We then consider a $(D-d)$-dimensional locally homogeneous space
with metric $\d s_{\text{Hom}}^2=\tilde{g}_{ab}(x^c)\d x^a\d x^b$; this space could
be an Einstein space such as, for example, $\mathbb{E}^{D-d}$,
$S^{D-d}$ or $\mathbb{H}^{D-d}$. We then take the product manifold
with metric
\begin{equation}
\d s_{\text{CSI}}^2 = \omega^2 \d s_{\text{VSI}}^2+\d s_{\text{Hom}}^2, \label{csir}
\end{equation}
where $H$ and $W_i$ are now possibly fibred (e.g., $H(v,u,x^k)$, $W_{i}(v,u,x^k))$ ($i$ and $k$ run possibly over all tranverse coordinates). If we take $\d s_{\text{Hom}}^2$ to be
Euclidean space, the Ricci tensor is of type $II$ (the Lorentzian
conformal-VSI part is of Weyl type $III$). These are CSI$_R$
spacetimes (belonging to the higher-dimensional Kundt CSI class,
CSI$_K$), and have been constructed in such a way as to be
solutions of supergravity. There will be solutions that preserve
supersymmetry. In particular, there is a subclass of these CSI$_R$
spacetimes which is also CCNV (i.e., the subclass with $w^2\equiv
1$ which is constructed from a CCNV VSI).

\section{Supergravity examples}
Let us provide some explicit examples of CSI supergravity
spacetimes. The examples illustrate a useful method of
constructing such spacetimes and, at the same time, are
interesting as possible solutions of higher-dimensional gravity
theories and supergravity. All of our examples are of the form of
metric (\ref{HKundt}) satisfying eqns.  (\ref{HKa}) and
(\ref{HKb}). The way these are constructed is as follows:
\emph{(i)} First we find a homogeneous spacetime,
$(\mathcal{M}_{\text{Hom}},\tilde{g})$, of Kundt form. Since there
is a wealth of such spacetimes we will concentrate on those that
are Einstein; i.e., that satisfy
$\widetilde{R}_{\mu\nu}=\lambda\widetilde{g}_{\mu\nu}$.
\emph{(ii)} We then generalise these spacetimes to include inhomogeneous
spacetimes, $(\mathcal{M},{g})$, by including arbitrary functions
${W}_{ i}^{(0)}(u,x^k)$, ${H}^{(1)}(u,x^k)$ and
${H}^{(0)}(u,x^k)$. By construction, the curvature invariants of
$(\mathcal{M},{g})$ will be identical to those of
$(\mathcal{M}_{\text{Hom}},\tilde{g})$. These spacetimes can, for
example, be used as the Lorentzian piece in the Freund-Rubin
construction.

Since the ``background'' homogeneous spacetime $(\mathcal{M}_{\text{Hom}},\tilde{g})$ is Einstein, these can be used as Freund-Rubin backgrounds, as explained. If we want to include matter, such as for example a scalar field, $\phi$, and a set of form-fields corresponding to a certain supergravity theory, the functions ${W}_{ i}^{(0)}(u,x^k)$, ${H}^{(1)}(u,x^k)$ and
${H}^{(0)}(u,x^k)$ will have to satisfy the corresponding supergravity equations involving the scalar field and form-fields. These form-fields will depend on the theory under consideration, and consequently also the corresponding field equations. In general, a form-field $F$, has the following boost-weight decomposition: 
\[ F=(F)_1+(F)_0+(F)_{-1}, \]
where $(F)_b$ denotes the projection onto the boost-weight $b$ components. For the Freud-Ruben solutions, the term $(F)_0$ is non-zero. These components typically imply that the curvatures of the Freund-Rubin background $M\times K$ are non-zero (hence, implying $\widetilde{R}_{\mu\nu}=\lambda\widetilde{g}_{\mu\nu}$ for  $M$). 

For the CSI spacetimes, we must demand that $(F)_1=0$. Therefore, the appropriate ansatz for the fields are 
\[   F=(F)_0+(F)_{-1}, \]
for the total space. In general, this ansatz will give boost-weight 0, $-1$ and $-2$ contributions to the supergravity equations (see \cite{CFHP} for details). 
Instead of solving the equations for each of the possible backgrounds and for each the possible matter fields (as in \cite{CFHP}, for which there was a managable number of cases), we will just provide a general construction how to find metrics of these types. These metrics are therefore supergravity solutions for a given set of fields, provided that the functions $W_i^{(0)}(u,x^k)$, ${H}^{(1)}(u,x^k)$ and
${H}^{(0)}(u,x^k)$ satisfy a set of differential equations. 

\subsection{$(\mathcal{M}_{\text{Hom}},\tilde{g})$ is a regular Lorentzian Einstein solvmanifold} \label{ExSol}
By using standard Einstein solvmanifolds, and Wick rotating, we
can get many examples of homogeneous Einstein Kundt metrics
\footnote{Note that in \cite{Her1} only a non-zero $H^{(0)}$ was considered. The metrics presented here are thus generalisations of those in \cite{Her1}.} \cite{Her1}. All of these spacetimes can be written as follows:
\beq 
g^{\perp}_{ij}(x^k)\d x^i\d x^j=\d w^2+\sum_{i}\exp(-2q_i
w)(\mbold\omega^i)^2, 
\eeq where $\{\mbold\omega^i\}$ is a
left-invariant metric of some subgroup\footnote{\;If the
solvmanifold is of rank one, this subgroup would be the nilpotent
group corresponding to the Einstein nilradical.}, \beq
{W}_{i}^{(1)}(u,x^k)\d x^i=2p\d w, \quad \tilde\sigma=0, \eeq and
$p=\sum_iq_i^2/(\sum_i q_i)$. The boost-weight decomposition of
$S$ (the trace-free Ricci tensor) and $C$ (the Weyl tensor) is as
follows:
\begin{itemize}
\item{} General ${W}_{i}^{(0)}(u,x^k)$, ${H}^{(1)}(u,x^k)$ and ${H}^{(0)}(u,x^k)$:
\[ S=(S)_{-1}+(S)_{-2}, \quad C=(C)_0+(C)_{-1}+(C)_{-2}.\]
\item{} ${W}_{i}^{(0)}(u,x^k)=0$, ${H}^{(1)}(u,x^k)=0$, general ${H}^{(0)}(u,x^k)$:
\[ S=(S)_{-2}, \quad C=(C)_0+(C)_{-2}.\]
\item{} An Einstein case:  ${W}_{i}^{(0)}(u,x^k)=0$, ${H}^{(1)}(u,x^k)=0$, and
\[ \Box^{\perp}H^{(0)}+\left(H^{(0)}W^{(1)}_i\right)^{;i}=0, \]
where $\Box^{\perp}$ is the Laplacian on the transverse space, and
\[ S=0, \quad C=(C)_0+(C)_{-2}\]
\end{itemize}
There is a cornucopia of examples of these metrics and
the simplest one corresponds to
$(\mathcal{M}_{\text{Hom}},\tilde{g})$ being AdS space (for which
$(C)_0=0$). The corresponding inhomogeneous Einstein metric with
$H^{(0)}\neq 0$ is the Siklos spacetime \cite{Siklos}.

There are a few special metrics in this class worth mentioning. A
special Siklos metric is the Kaigorodov spacetime  \cite{kaigor}
which is both Einstein and homogeneous (see section $3.3$).
Another special homogeneous metric is the conformally flat metric:
\beq \d s^2=2e^{-2qz}\d u\left(\d v+ae^{qz}\d u\right)+e^{-2qz}\d
y^2+\d z^2. \eeq This metric has vanishing Weyl tensor, $C=0$,
while $S=(S)_{-2}$.\footnote{In addition to a (negative) cosmological constant this metric can be sourced by, for example, an electromagnetic field of the form $F=2q\sqrt{a}\exp(-qz/2)\d u\wedge\d z$.} Both this metric, and the Kaigorodov metric,
are homogeneous Kundt metrics having identical curvature
invariants to AdS.

There are many 'non-trivial' examples of this type as well. As an illustration, the following Kundt metric is a 7-dimensional regular Lorentzian Einstein solvmanifold:
\beq
\d s^2&=&2\d u\left(\d v+3pv\d r\right)+e^{-4pr}(\d x-y\d w)^2+e^{-3pr}(\d y-z\d w)^2\nonumber \\
&& +e^{-2pr}\d z^2+e^{-pr}\d w^2+\d r^2,
\eeq
where $p=1/(2\sqrt{2})$. This metric has $\widetilde{R}_{\mu\nu}=-(3/2)\widetilde{g}_{\mu\nu}$ and can be generalised to the inhomogeneous case by the standard procedure.
\subsection{Some 5D examples}
Let us consider some non-trivial examples which can \emph{not} be obtained by a Wick-rotation of an Einstein solvmanifold. Therefore, these are not contained in \cite{Her1} and are believed to be new. The general construction of Einstein metrics of this kind is given in the Appendix. 

\subsubsection{Transverse space is the Heisenberg group} \label{subsubHeis3}
The transverse space is the Heisenberg group with a left-invariant metric:
\[ g^{\perp}_{ij}(x^k)\d x^i\d x^j=\left(\d x+\frac b2(y\d z-z\d y)\right)^2+\d y^2+\d z^2, \]
and
\[ W^{(1)}_i\d x^i=\sqrt{2}b\left(\d x+\frac b2(y\d z-z\d y)\right), \quad \tilde\sigma=\frac{b^2}{4}.\]
Here, $\widetilde{R}_{\mu\nu}=-(b^2/2)\widetilde{g}_{\mu\nu}$.
The Weyl tensor decomposes as
\[ C=(C)_0+(C)_{-1}+(C)_{-2}.\]
For the trace-free Ricci tensor:
\begin{itemize}
\item{} General ${W}_{i}^{(0)}(u,x^k)$, ${H}^{(1)}(u,x^k)$ and ${H}^{(0)}(u,x^k)$:
$ S=(S)_{-1}+(S)_{-2}$.
\item{} ${W}_{i}^{(0)}(u,x^k)=0$, ${H}^{(1)}(u,x^k)=0$, general ${H}^{(0)}(u,x^k)$:
$ S=(S)_{-2}$.
\item{} An Einstein case:  ${W}_{i}^{(0)}(u,x^k)=0$, ${H}^{(1)}(u,x^k)=0$, and
\[ \Box^{\perp}H^{(0)}+\left(H^{(0)}W^{(1)}_i\right)^{;i}=0, \]
where $\Box^{\perp}$ is the Laplacian on the transverse space.  Given that $H^{(0)}(u,x^k)$ satisfies this equation, this is an Einstein space, and hence $S=0$. The general solution to this equation can be found using standard methods (for example, separation of variables).
\end{itemize}
\subsubsection{Transverse space is $SL(2,\mathbb{R})$}
The transverse space is  $SL(2,\mathbb{R})$ with a left-invariant metric:
\[ g^{\perp}_{ij}(x^k)\d x^i\d x^j=\left(\d x-a\frac{\d z}{y}\right)^2+\frac{b^2}{y^2}(\d y^2+\d z^2),\]
and
\[  W^{(1)}_i\d x^i=\frac{\sqrt{2(a^2+b^2)}}{b^2}\left(\d x-a\frac{\d z}{y}\right), \quad \tilde\sigma=\frac{a^2}{4b^4}.\]
Here, $\widetilde{R}_{\mu\nu}=-[(a^2+2b^2)/(2b^4)]\widetilde{g}_{\mu\nu}$.
The Weyl tensor decomposes as
\[ C=(C)_0+(C)_{-1}+(C)_{-2}.\]
For the trace-free Ricci tensor:
\begin{itemize}
\item{} General ${W}_{i}^{(0)}(u,x^k)$, ${H}^{(1)}(u,x^k)$ and ${H}^{(0)}(u,x^k)$:
$ S=(S)_{-1}+(S)_{-2}$.
\item{} ${W}_{i}^{(0)}(u,x^k)=0$, ${H}^{(1)}(u,x^k)=0$, general ${H}^{(0)}(u,x^k)$:
$ S=(S)_{-2}$.
\item{} An Einstein case:  ${W}_{i}^{(0)}(u,x^k)=0$, ${H}^{(1)}(u,x^k)=0$, and
\[ \Box^{\perp}H^{(0)}+\left(H^{(0)}W^{(1)}_i\right)^{;i}=0, \]
where $\Box^{\perp}$ is the Laplacian on the transverse space. Given that $H^{(0)}(u,x^k)$ satisfies this equation, this is an Einstein space, and thus $S=0$. The general solution to this equation can be found using standard methods (for example, separation of variables).
\end{itemize}

\subsubsection{Transverse space is the 3-sphere, $S^3$} \label{ExS3}
The transverse space is the 3-sphere, $S^3$, with the Berger metric:
\[ g^{\perp}_{ij}(x^k)\d x^i\d x^j=a^2\left(\d x+\sin y{\d z}\right)^2+{b^2}(\d y^2+\cos^2 y\d z^2), \]
and
\[  W^{(1)}_i\d x^i=\frac{a\sqrt{2(a^2-b^2)}}{b^2}\left(\d x+\sin y{\d z}\right), \quad \tilde\sigma=\frac{a^2}{4b^4}.\]
Here, $\widetilde{R}_{\mu\nu}=-[(a^2-2b^2)/(2b^4)]\widetilde{g}_{\mu\nu}$, and hence, can be positive, zero or negative.
The Weyl tensor always decomposes as
\[ C=(C)_0+(C)_{-1}+(C)_{-2}.\]
For the trace-free Ricci tensor:
\begin{itemize}
\item{} General ${W}_{i}^{(0)}(u,x^k)$, ${H}^{(1)}(u,x^k)$ and ${H}^{(0)}(u,x^k)$:
$ S=(S)_{-1}+(S)_{-2}$.
\item{} ${W}_{i}^{(0)}(u,x^k)=0$, ${H}^{(1)}(u,x^k)=0$, general ${H}^{(0)}(u,x^k)$:
$ S=(S)_{-2}$.
\item{} An Einstein case:  ${W}_{i}^{(0)}(u,x^k)=0$, ${H}^{(1)}(u,x^k)=0$, and
\[ \Box^{\perp}H^{(0)}+\left(H^{(0)}W^{(1)}_i\right)^{;i}=0, \]
where $\Box^{\perp}$ is the Laplacian on the transverse space. Given that $H^{(0)}(u,x^k)$ satisfies this equation, this is an Einstein space, and thus $S=0$.
\end{itemize}

\subsection{Examples in the literature}
A number of special cases of the examples discussed in the
previous
two subsections are known, and the supersymmetry properties of many of them
have been discussed. All of the examples given below are in the subclass of $CSI_R$ spacetimes. Let us review these examples briefly. \\

We give in the first place an example of a CCNV CSI. In
\cite{cnmcsi} the following five-dimensional metric was
considered: 
\be \d s^2=2\d u\left[\d v +K(u,x^k)\;\d u\right]+\d
\xi^2 + \sin^2{\xi} \d \theta^2 + \sin^2{\xi} \sin^2{\theta} \d
\phi^2  \label{3sph} \ee 
The transverse space is $S^3$ with unit
radius and the function $K$ satisfies 
\be \Box^{\perp}K=0
\label{KS^3} 
\ee 
where $\Box^{\perp}$ is the Laplacian on $S^3$.
The covariantly constant null Killing vector is $\partial_v$. Note
that metric (\ref{3sph}) is already in the Kundt form
(\ref{HKundt}), with ${W}_{i}^{(1)}={W}_{
i}^{(0)}=\tilde{\sigma}={H}^{(1)}=0$.
The metric (\ref{3sph}), together with a constant dilaton and appropiate
antisymmetric field, is an exact solution to bosonic string theory\footnote{\;However, it is not a vacuum solution of five-dimensional gravity.}. \\

The next two examples are not CCNV, but are constructed from a
CCNV VSI (see section $2.2$). As such they have the null Killing
vector $\partial_{v}$; however, this vector is no longer
covariantly constant due to the introduction of a warp factor.
Recall that if the VSI seed metric is Ricci flat they are Einstein
spaces. The first example is the $d$-dimensional Siklos spacetime
\be 
ds^2 = \frac{l^2}{z^2} \left[ 2\d u\d v + 2H(u,x^k)\;\d u^2 +
(\d x^i)^2 +\d z^2\right] \label{siklos}, 
\ee where $i=1,\ldots,d-3$. The
Siklos metric can be cast in the Kundt form (\ref{HKundt}) by
making a coordinate transformation $\tilde{v}=vl^2/z^2$ 
\be 
\d s^2=2\d u\left(\d \tilde{v} + \frac{l^2}{z^2}H(u,x^k)\;\d u +
\frac{2\tilde{v}}{z}\;\d z\right) + \frac{l^2}{z^2}\left[(\d x^i)^2+\d z^2\right]
\label{sikloskundt} \ee In this way
$\tilde{\sigma}={H}^{(1)}={W}_{i}^{(0)}=0$,
$H^{(0)}=(l^2/z^2)H(u,x^k)$ and ${W}_{z}^{(1)}=2/z$; the
transverse space is $\mathbb{H}^{d-2}$. In the new coordinates the
null Killing vector is $l^2/z^2\partial_{\tilde{v}}$. The
Kaigorodov metric $K_d$ is a Siklos spacetime with $H=z^{d-1}$
\cite{kaigor,cvetic}. 
Since it is homogeneous, it has at least $d$ Killing vectors
(but only $\partial_{v}$ can be null). 
The Siklos spacetime is of Weyl type $N$. \\

All of the Siklos metrics preserve $1/4$ of the supersymmetries, regardless the form of the function $H$ in (\ref{siklos}) \cite{Brecher}. This was previously shown for the Kaigorodov metric in \cite{cvetic}. \\

The second example is the $d$-dimensional $AdS$ gyraton, with metric \cite{FZ}
\begin{equation}
\d s^2 =  \frac{l^2}{z^2}  \left[ 2\d u\d v + 2H(u,x^k)\d u^2
+2W_{i}(u,x^k)\d u\d x^i + (\d x^i)^2 +\d z^2\right], \label{adsgyr}
\end{equation}
where $i=1,\ldots,d-3$ and $H$ and $W_i$ are independent of $v$.
In the Kundt form we have (\ref{sikloskundt}) but additionally
${W}_{ i}^{(0)}=\frac{l^2}{z^2}W_{i}$; the null Killing vector is
$l^2/z^2\partial_{\tilde{v}}$ as for the Siklos metric. This is a
metric of the form given in section \ref{ExSol} where the
homogeneous space is $AdS_d$. The Weyl type is $III$. The
five-dimensional $AdS$ gyraton has been considered in the context
of gauged supergravity, and both gauged and ungauged supergravity
coupled to an arbitrary number of vector supermultiplets
\cite{Caldarelli}. Some of
these solutions preserve $1/4$ of the supersymmetry \cite{Caldarelli,Gutowski:2005}. \\

We consider now metrics of the form (\ref{csir}). The most
well-known examples in this class are the $AdS \times S$ spaces.
Let us discuss $AdS_5 \times S^5$ \be \d s^2=\frac{1}{z^2} \left[
2\d u\d v + \d x^2 + \d y^2 + \d z^2 \right]+\d \Omega^2_5
\label{adss5} \ee where $\d \Omega^2_5$ is the standard round
metric on the unit\footnote{\;We can multiply (\ref{adss5}) by
$l^2$; then $r^2=1/l^2$ is the radius of $S^5$.} $5$-sphere. This
is clearly of the form (\ref{csir}), with the simplest VSI
(Minkowski) spacetime. It is a (maximally symmetric) Einstein
space. In the Kundt form (\ref{HKundt}) \be \d s^2=2\d u\left(\d
\tilde{v} + \frac{2\tilde{v}}{z}\d z\right) +
\frac{1}{z^2}\left[\d x^2 + \d y^2 + \d z^2 \right]+\d \Omega^2_5
\ee with $\tilde{\sigma}={H}^{(1)}=H^{(0)}={W}_{i}^{(0)}=0$,
${W}_{z}^{(1)}=2/z$; the transverse space is
$\mathbb{H}^3 \times S^5$. It is of Weyl type $O$ (provided their sectional curvatures have equal magnitude and opposite sign, otherwise they are Weyl type D).

Spaces of the form $AdS \times S$, together with appropriate five- or four-form
fields, are maximally supersymmetric solutions of IIB and eleven-dimensional supergravities
\cite{Freund, Pilch, Schwarz}. \\

$AdS_5 \times S^5$ can be generalized by considering other VSI
seeds. The resulting metrics are of Weyl type $III$ at
most\footnote{\;These spacetimes are of type II if the sectional curvatures
 are not of equal magnitude and opposite sign.}. For example, \be \d s^2=\frac{1}{z^2} \left[
2\d u\d v + 2H(u,x,y,z,x^a)\d u^2 + \d x^2 + \d y^2 + \d z^2
\right]+\d \Omega^2_5 \ee where $x^a$ are the coordinates on
$S^5$. In the Kundt form we have now $H^{(0)}=H/z^2$. Such
spacetimes are supersymmetric solutions of IIB supergravity (and
there are analogous solutions in $D=11$ supergravity)
\cite{kumar}. Supersymmetric solutions of this type in
$D=5$ gauged supergravity were given in \cite{kerimo}, where $\d s_{\text{Hom}}^2$
was taken to be flat (Weyl type $N$).   \\

The idea of considering spaces of the form $AdS \times M$, with
$M$ an Einstein (-Sasaki) manifold other than $S^n$, goes back to
\cite{membranes}. Such spaces have Weyl type $II$. In
\cite{membranes} supersymmetric solutions of $D=11$ supergravity
of Weyl type $II$ are presented where, for example, $M$ is the
squashed $S^7$. Examples where $M$ is taken to
flat and hyperbolic space can be found in \cite{cardoso} (in the context of higher-dimensional
Einstein-Maxwell theory). In ten dimensions, solutions of the form $AdS_5 \times T^{1,1}$
have been extensively studied. Recently, an infinite class of five-dimensional Einstein-Sasaki
spaces (called $Y^{p,q}$) has attracted much attention\footnote{\;However, these are not homogeneous and hence, not CSI. There are many homogeneous
Einstein spaces on $S^2\times S^3$ \cite{ADF}; however, $T^{1,1}$ is the only one that is also Sasaki.} \cite{sas-eins}.  \\

The final example concerns a warped product of $AdS_3$ with an
$8$-dimensional compact (Einstein-Kahler) space $M_8$:
\begin{equation}
\d s^2 = \omega^2 [ \d s^2(AdS_3) + \d s^2(M_8)].
\end{equation}
These metrics with non-vanishing 4-form flux are supersymmetric
solutions of D=11 supergravity \cite{Gauntlett}. Similar
constructions can be found in \cite{gaunt1}.

\section{Conclusion}

In this paper we have discussed a (sub)class of CSI spacetimes
which are solutions of supergravity. We have utilized two
different approaches. In the {\em top-down} approach we considered
a subclass of known CSI product manifolds of the form $M\times K$
and investigated the conditions under which they will be solutions
of supergravity. In a {\em bottom-up} approach we built CSI Kundt
spacetimes using a Lorentzian VSI spacetime and a known
homogeneous spacetime as seeds \cite{CSI}, which by construction
will automatically be solutions of supergravity. We also discussed
which of these CSI supergravity solutions may preserve
supersymmetries.

We have explicitly constructed a number of new exact CSI
supergravity solutions, some of which are generalizations of
$AdS\times S$ spacetimes and $AdS$ gyratons. Indeed, in some of
the simple generalizations of $AdS\times S$ spacetimes all of the
curvature invariants are identical to those of $AdS\times S$,
which may be of importance when considering higher order
corrections \cite{final} 
(i.e., it is plausible that these generalizations are also
exact string solutions). 
The newly constructed spacetimes include
solutions that are based on (warped) products of regular
Lorentzian Einstein solvmanifolds (including the simple Siklos
metric) and transverse spaces which are ($D-d$)-spheres (as well
as squashed spheres and Euclidean and hyperbolic spaces). 
Finally, we have reviewed the known CSI supergravity solutions, and 
we have shown that they belong to the higher-dimensional Kundt class.

\section*{Acknowledgements} This work was supported
by NSERC (AC),  AARMS (SH) and the
programme FP52 of the Foundation for Research of Matter, FOM (AF).

\appendix

\section{Constructing homogeneous Einstein Kundt metrics}

Let us briefly discuss the general method for constructing the
homogeneous Kundt metrics illustrated in subsections (3.1) and
(3.2) (the examples given in 5D are easily generalized to higher
dimensions). Consider a Lie group $G$ equipped with a
left-invariant frame ${\bf m}^i$. A class of Lorentzian Kundt
metrics can then be written: 
A set of
\beq \d s^2=2\d u\left(\d
v+v^2\tilde{\sigma} \d u+v\beta_i{\bf m}^i\right)+\delta_{ij}{\bf
m}^i{\bf m}^j, \eeq where $\tilde{\sigma}$ and $\beta_i$ are
constants. This is automatically a homogeneous space with
left-invariant frame  
\beq {\mbold\omega^0}=v\d u, \quad
{\mbold\omega^1}=\frac{\d v}{v}+v\tilde\sigma\d u+\beta_i{\bf
m}^i, \quad {\mbold\omega}^{i+1}={\bf m}^i. 
\eeq 
A set of
transitively acting Killing vectors are:
\[ {\mbold\xi}_0=v\frac{\partial}{\partial v}-u\frac{\partial}{\partial u}, \quad {\mbold\xi}_1=\frac{\partial}{\partial u}, \quad {\mbold\xi}_{i+1}={\mbold\xi}^G_i,\]
where ${\mbold\xi}^G_i$ is a set of transitively acting Killing vectors on $G$.\footnote{This homogeneous Kundt metric can therefore be considered as the the Lie group $A_2\times G$, where $A_2$ is the unique 2-dimensional non-abelian Lie group.} 

Let $\widetilde{R}_{ij}$ be the Ricci tensor of $\delta_{ij}{\bf m}^i{\bf m}^j$. Then
\beq
R_{01}&=& \frac 12(4\tilde{\sigma}+C^j_{~ji}\beta^i-\beta_i\beta^i), \\
R_{ij}&=& \widetilde{R}_{ij}+C_{(ij)k}\beta^k-\frac
12\beta_i\beta_j, \eeq where $\beta_{(i;j)}=C_{(ij)k}\beta^k$ and
$C^i_{~jk}$ are the structure constants of the Lie group $G$. We
have not written down the Ricci components of boost-weights $-1$
and $-2$.

The examples essentially split into two different cases according
to whether $G$ is unimodular or not. The regular Lorentzian
solvmanifolds are not unimodular and can be found using the
Riemannian analysis. The unimodular case, $C^i_{ij}=0$,
corresponding to true 'authentic' Lorentzian solutions and have to
be found on a case-by-case basis (such solutions were not considered in \cite{Her1}).

\subsection{All 5D homogeneous Einstein manifolds of this type}
It can be shown that all 5D manifolds of this type with
($\beta_i\neq 0$) are given in the text or are Lorentzian versions
of standard Einstein solvmanifolds. The classification of 3D Lie
algebras is well-known and are enumerated I-IX using the Bianchi
classification. The Lie algebras of the Heisenberg group,
$SU(2)\cong S^3$ and $SL(2,\mathbb{R})$ are $II$, IX and VIII,
respectively. The above method for the other Lie algebras also
gives an Einstein metric for the type III algebra; however, the
metric is the same as the $SL(2,\mb{R})$ since this also admits a
simply transitive type III action.

\subsection{Other examples}

There are a few other examples.
\begin{itemize}
\item{} $G$ is the $(2m+1)$-dimensional Heisenberg group. The spacetime is of dimension $(3+2m)$ and metric is similar to the $m=1$ case in subsection \ref{subsubHeis3}.
\item{} $G$ is $S^3\times S^3$: Given the left-invariant one-forms $\sigma^i$ and $\hat{\sigma}^i$ on the two $S^3$ so that
\[
\d \sigma^i=\frac 12\varepsilon^i_{~jk}\sigma^j\wedge\sigma^k, \quad \d \hat\sigma^i=\frac 12\varepsilon^i_{~jk}\hat\sigma^j\wedge\hat\sigma^k,
\]
the metic can be written:
\beq
\d s^2&=&2\d u\left[\d v+\tilde\sigma v^2\d u+\alpha(\sigma^1+\hat\sigma^1)\right]\nonumber \\
&&+A^2\left[(\sigma^1)^2+2\lambda\sigma^1\hat\sigma^1+(\hat\sigma^1)^2\right]+B^2\left[(\sigma^2)^2+(\sigma^3)^2+(\hat\sigma^2)^2+(\hat\sigma^3)^2\right],
\eeq
where $A^2/2< B^2\leq A^2$ and
\[ \lambda=\frac{2(A^2-B^2)}{A^2}, \quad \tilde\sigma=\frac{3A^2-2B^2}{2B^4}, \quad \alpha^2=\frac{2(A^2-B^2)(3A^2-2B^2)}{B^4}.\]
This is a positively curved Einstein space.
\end{itemize}


\begin{thebibliography}{99}


\bibitem{VSISUG} A. Coley, A. Fuster, S. Hervik and N. Pelavas, JHEP {\bf 0705}, 032 (2007). 


\bibitem{ss1}
D. Amati and C. Klim\v c\'\i k, Phys. Lett. B {\bf 219}, 443
(1989); G.T. Horowitz and A.R. Steif, Phys. Rev. Lett. {\bf 64}
260 (1990); A.A. Coley, Phys. Rev. Lett. {\bf 89}, 281601 (2002).


\bibitem{ss2}
R.~R.~Metsaev and A.~A.~Tseytlin, Phys. Rev.  D {\bf 65}, 126004
(2002); M. Blau et al., JHEP {\bf 0201}, 047 (2002); P. Meessen,
Phys. Rev. D {\bf 65}, 087501 (2002); J. G. Russo and A.A.
Tseytlin, JHEP {\bf 0209}, 035 (2002); J.~Maldacena and L.~Maoz,
JHEP {\bf 0212}, 046 (2002).


\bibitem{CSI}    A. Coley, S. Hervik and N.
Pelavas, Class. Quant. Grav. {\bf 23}, 3053 (2006).


\bibitem{Gauntlett} J. Gauntlett et al., Phys. Rev. D {\bf 74}, 106007 (2006). 

\bibitem{hortseyt} G. T. Horowitz  and A. A. Tseytlin, Phys. Rev. D {\bf 51}, 2896
(1995).

\bibitem{Caldarelli} M. Caldarelli et al., Class. Quant. Grav. {\bf 24}, 1341 (2007).  

\bibitem{FZ} V. P. Frolov and A. Zelnikov, Phys. Rev. D {\bf 72},
104005 (2005); V. P. Frolov and D. V. Fursaev, Phys. Rev. D {\bf
71}, 104034 (2005).


\bibitem{Kallosh} R. Kallosh et al., Phys. Rev. D {\bf 58}, 125003 (1998). 


\bibitem{class}  A. Coley, R. Milson, V. Pravda and A. Pravdova,
Class. Quant. Grav. {\bf 21}, L35 (2004).

\bibitem{3dCSI} A. Coley, S. Hervik and N. Pelavas, Class. Quant. Grav. {\bf 25}, 025008 (2008).

\bibitem{CFHP} A. Coley, A. Fuster, S. Hervik and N. Pelavas, Class. Quant. Grav. {\bf 23}, 7431 (2006). 




\bibitem{Higher} A. Coley, R. Milson, V. Pravda and A. Pravdova,
Class. Quant. Grav. {\bf 21}, 5519 (2004).


\bibitem{Acharya} B. S. Acharya, J. M. Figueroa-O'Farrill, C. M. Hull and B.
Spence, Adv. Theor. Math. Phys. {\bf 2} 1249 (1999).


\bibitem{Guven}  R. Guven, Class. Quant. Grav. {\bf 23}, 295 (2006).


\bibitem{Her1} S. Hervik, J. Geom. Phys. {\bf 52}, 298 (2004); S.Hervik, Class. Quant. Grav. {\bf 21}, 4273 (2004).


\bibitem{Siklos}
S.T.C. Siklos, Lobatchevski plane gravitational waves, in
\textit{Galaxies, axisymmetric systems and relativity} ed. M.A.H.
MacCallum, Cambridge University Press, 1985.


\bibitem{kaigor}
V. Kaigorodov, Dokl. Akad. Nauk. SSSR {\bf 146}, 793 (1962); Sov.
Phys. Doklady {\bf 7}, 893 (1963).


\bibitem{cnmcsi} G.T. Horowitz and A.A. Tseytlin, Phys. Rev. D {\bf 50}, 5204 (1994).

\bibitem{Brecher}
D. Brecher, A. Chamblin and H. S. Reall, Nucl. Phys. B {\bf 607}, 155 (2001).


\bibitem{cvetic}
M. Cvetic, H. Lu and C. N. Pope, Nucl. Phys. B {\bf 545}, 309 (1999).


\bibitem{Gutowski:2005}
J. B. Gutowski and W. Sabra, JHEP {\bf 10}, 039 (2005); J. P.
Gauntlett and J. B. Gutowski, Phys. Rev. D {\bf 68}, 105009 (2003).


\bibitem{Freund}
P. G. O. Freund and M. A. Rubin, Phys. Lett. B {\bf 97}, 233
(1980).

\bibitem{Pilch}
K. Pilch, P. van Nieuwenhuizen and P. K. Townsend, Nucl. Phys. B
{\bf 242}, 377 (1984).

\bibitem{Schwarz}
J. H. Schwarz, Nucl. Phys. B {\bf 226}, 269 (1983).



\bibitem{kumar} A. Kumar and H. K. Kunduri, Phys. Rev. D {\bf 70}, 104006 (2004).

\bibitem{kerimo} J. Kerimo, JHEP {\bf 0509}, 025 (2005).


\bibitem{membranes}
M.J. Duff, H. Lu, C.N. Pope and E. Sezgin, Phys. Lett. B {\bf 
371}, 206 (1996).

\bibitem{cardoso} V. Cardoso, O. J. C. Dias and J. P. S. Lemos,
Phys. Rev. D {\bf 70}, 024002 (2004).


\bibitem{ADF}
D. Alekseevsky, I. Dotti and C. Ferraris, Pacific J.
Math. \textbf{175}, 1 (1996).



\bibitem{sas-eins}
J. P. Gauntlett, D. Martelli, J. Sparks and D. Waldram, Adv.
Theor. Math. Phys. {\bf 8}, 711 (2004).



\bibitem{gaunt1}
J. P. Gauntlett, D. Martelli, J. Sparks and D. Waldram, Class.
Quant. Grav. {\bf 21}, 4335 (2004) \& {\bf 23}, 4693 (2006).



\bibitem{final} P. Meessen, arXiv:0705.1966.









\end{thebibliography}
\end{document}